\documentclass[twocolumn,showpacs,amsmath,amssymb,pre,superscriptaddress]{revtex4}
\usepackage{graphicx,color,longtable}

\begin{document}
\title{Incompressibility of polydisperse random close packed
colloidal particles}
\author{Rei Kurita} 
\affiliation{Institute of Industrial Science, 
The University of Tokyo, 4-6-1 Komaba, Meguro-ku, Tokyo 153-8505, Japan}
\author{Eric R. Weeks}
\affiliation{Department of Physics, Emory University, 
Atlanta, Georgia 30322, USA}

\date{\today}

\begin{abstract}
We use confocal microscopy to study a random close packed sample
of colloidal particles.  We introduce an algorithm to estimate
the size of each particle.  Taking into account their sizes, we
compute the compressibility of the sample as a function of wave
vector $q$, and find that this compressibility vanishes linearly
as $q \rightarrow 0$.  The particle sizes must be considered to
calculate the compressibility properly.  These results also
suggest that the experimental packing is hyperuniform.
\end{abstract}
\maketitle

The random packing of objects has been studied scientifically
for nearly a century \cite{smith29,westman30}; see
Ref.~\cite{torquato10} for a review.  This problem is often
termed ``random close packing'' (rcp) or ``maximally random
jammed packing'' \cite{Torquato2000}.  Important recent work has
focused on the packing of highly polydisperse systems
\cite{clusel09}, ellipsoids \cite{donev04}, and tetrahedra
\cite{chenglotzer10}, but the simplest packing
problem is the packing of monodisperse spheres.  In the
past decade, simulations studying monodisperse spheres have
generated large rcp configurations with $10^5 - 10^6$ spheres
\cite{Torquato2005,Silbert}.  These simulations enable study of
density fluctuations at very large length scales, or equivalently,
small wave vectors $q$.  They find that the static structure
factor $S(q)$ approaches zero linearly as $q \rightarrow 0$, that
is, $S(q) \sim q$ for small $q$.  This finding has been termed
``hyperuniformity'' \cite{Torquato2005}.  One corollary is that
the sample is incompressible, as the isothermal compressibility
$\chi$ in simple liquids can be found from $\rho k_BT \chi
= S(0)$ where $\rho$, $k_B$, and $T$ are the mean density,
Boltzmann constant, and temperature.  These observations of
close-packed samples are in contrast, for example, with simple
liquids for which $S(0) > 0$ \cite{Berthier}.  The existence
of hyperuniformity has been seen in a variety of systems, see
for example discussions in Refs.~\cite{torquato10,Zachary1}.
In general, long wavelength density fluctuations are important
for diverse fields including critical phenomena \cite{Onuki} and
the shear flow of glassy materials \cite{Furukawa}.  Likewise,
understanding random close packed samples is relevant for
understanding liquids, glasses, biological systems, and granular
materials \cite{smith29,Bernal,torquato10}.

In 2010 we published an experimental study of a random close
packed sample of colloidal particles, observed with confocal
microscopy \cite{KW}.  Our data set was the positions of more than
500 000 slightly polydisperse particles \cite{epaps}, and we found that $S(q
\rightarrow 0) > 0$, implying that the experimental sample was
compressible and not hyperuniform.  A 2010 simulation of a binary
sample found similar results \cite{Xu2}.  These results seem to
demonstrate random close packed samples that are not hyperuniform.
However, in 2011 two groups showed that in polydisperse samples,
careful consideration of the individual particle sizes recovers
hyperuniformity and incompressibility \cite{Berthier,Zachary1}.
In particular, Berthier {\it et al.} showed how to compute the
isothermal compressibility when the individual particle sizes are
known, and demonstrated that samples with $S(0) > 0$ nonetheless
can be incompressible \cite{Berthier}.  They examined data from a
two-dimensional granular experiment and confirmed that $\chi(0)
= 0$.  The reason $S(0)>0$ in polydisperse systems is because
density fluctuations are coupled to composition fluctuations,
but such samples can still be incompressible and hyperuniform.

In this article, we describe a method to determine each particle
size from microscopy observations of a random close packed sample
of colloidal particles.  We use numerically generated packings to
confirm that our method accurately determines the particle radii.
Analyzing our experimental data using the method of Berthier {\it
et al.} \cite{Berthier}, we confirm that our experimental system
is hyperuniform and incompressible.  We additionally note an
anticorrelation between the local polydispersity and local ordering.

As we use the analytical method introduced by Berthier {\it et
al} \cite{Berthier}, we briefly summarize their method here.
They consider a wave vector dependent isothermal compressibility
$\chi(q)$ which is related to the structure factor of a {\it
monodisperse} sample by $\rho k_B T \chi(q) = S(q)$.  They then
derive an exact formula relating $\chi(q)$ and $S(q)$ for a
polydisperse sample, although the formula is ``conceptually and
computationally difficult'' to evaluate \cite{Berthier}.  Thus,
they derive a series of approximate formulas, of which the first
order approximation is sufficient for samples of low polydispersity
such as ours.  To start with, they define single-particle density
fields $\rho_i(\boldsymbol{q}) = \exp (i \boldsymbol{q} \cdot
\boldsymbol{r_i})$ where $\boldsymbol{r_i}$ is the position of
particle $i$.  They also define the size deviation of particle
$i$ as $\epsilon_i = (a_i - \bar{a})/\bar{a}$, where $a_i$ is
the radius of particle $i$ and $\bar{a}$ is the mean radius.
(Note that $\sqrt{\langle \epsilon_i^2 \rangle} = p$ defines the
polydispersity $p$ of a sample.)  These $\epsilon_i$'s are the small
parameters used in the approximation.  Using these variables, they
define a 2 $\times$ 2 matrix $\boldsymbol{S}(q)$ with elements
$S^{uv}(q) = \frac{1}{N} \langle \epsilon^u(\boldsymbol{q})
\epsilon^v(-\boldsymbol{q}) \rangle$, with $u,v  \in  {0, 1}$,
$\epsilon^u(q) = \Sigma_{i=1}^N \epsilon_i^u \rho_i (q)$, and
$N$ is the total number of particles.  The matrix elements can
be used to provide a first order approximation $\chi_{1}(q)$
as $\rho k_B T \chi_{1}(q) = S^{00} - [S^{01}]^2/S^{11}.$
They confirm that $\chi_1(0) \approx 0$ in cases for which the
sample polydispersity is less than 10\%, while $S(0) \neq 0$
for those cases.  Their results suggest that random close packed
systems are hyperuniform and incompressible even when the sample
is polydisperse \cite{Berthier}.

In our prior work, we used colloidal particles to generate a random
close packed sample, and imaged this with confocal microscopy.
We reprise the key experimental points here; a more detailed
experimental discussion is in Ref.~\cite{KW}.  We use sterically
stabilized poly(methy methacrylate) (PMMA) particles \cite{Antl}
with $\bar{a}$ = 1.265 $\mu$m.  Previously we reported that these
particles had a polydispersity of $\sim 5$\% \cite{KW}; below, we
determine that the true polydispersity is 6.7\%.  The PMMA particles
are suspended in a solvent mixture that is slightly lower density
than the particles.  The sample is mixed and then the particles
are allowed to sediment until they are close packed.
We use a confocal microscope to take clear images deep inside our
dense sample \cite{Dinsmore}.  Overlapping images are taken, with
total volume 492 $\times$ 514 $\times$ 28 $\mu$m$^3$.  Within this
volume, particles are identified within 0.03 $\mu$m in $x$ and
$y$, and within 0.05 $\mu$m in $z$ \cite{Dinsmore, Crocker}.
The total data set contains 543 136 particles \cite{epaps}.

The average particle size $\bar{a}$ is obtained from the position
of the first peak of the pair correlation function \cite{KW}.
It is difficult to determine subtle size differences
between individual particles from microscopy due to diffraction.
However, obtaining the positions of each particle can be done
fairly accurately.  A large particle will be slightly farther
from its neighbors as compared to a small particle, and we use
this idea as a starting point for an estimation method for each
particle size.

Given that our sample is jammed, each particle must be in contact
with several of its neighbors.  In fact, a numerical simulation
of random close packed monodisperse particles showed that each
particle contacts with at least 6 particles \cite{Torquato2003}.
When particle $i$ contacts with particle $j$, the separation
between these two particles is given by $r_{ij} = a_i +
a_j$, where $a_i$ and $a_j$ are their radii.
The average of $r_{ij}$ over all neighbors $j$ is given by
$\langle r_{ij} \rangle_{j} = a_i + \langle
a_j \rangle_{j}$.
Next, consider separations $r_{jk}$ between particle
$i$'s contacting neighbors $j$ and contacting neighbors $k$ of those
particles.
Again, we take an
average of $r_{jk}$ with respect to particles $j$ and $k$,
giving $\langle \langle r_{jk} \rangle_{k}\rangle_{j} = \langle
a_j \rangle_{j} + \langle \langle a_k \rangle_{k} \rangle_{j}$.
Then we subtract $\langle \langle r_{jk} \rangle_{k}\rangle_{j}$
from $\langle r_{ij} \rangle_{j}$, leading to
\begin{equation}
a_i = \langle \langle a_k \rangle_{k} \rangle_{j} + 
\langle r_{ij} \rangle_{j} - \langle \langle r_{jk} \rangle_{k}\rangle_{j}.
\label{eq:2}
\end{equation}

We choose the 5 nearest particles from
particle $i$ as the particles $j$, assumed to be in contact with
particle $i$, and likewise for each particle $j$ we identify its
five closest neighbors for the particles $k$.
For each particle $j$, one of its neighbors $k$ 
should be particle $i$, leading to an overcounting in the
average:
$\langle \langle a_k \rangle_k\rangle_j = (1/5)a_i+(4/5)\langle \langle a_k
\rangle_{k \neq i}\rangle_j$.
Likewise, $\langle \langle r_{jk}
\rangle_{k}\rangle_{j} = (1/5) \langle r_{ij} \rangle_j + (4/5) \langle
\langle r_{jk} \rangle_{k \neq i}\rangle_{j}$ 
from the same overcounting of particle $i$.  Using these
results, we obtain
\begin{equation}
a_i = \langle \langle a_k \rangle_{k \neq i}\rangle_j +
      \langle r_{ij} \rangle_{j} -
      \langle \langle r_{jk} \rangle_{k \neq i}\rangle_{j}.
\label{Rioriginal}
\end{equation}
To compute $\langle \langle a_k \rangle_{k \neq i}\rangle_j$,
we use $\bar{a}$ as an initial guess for the particle sizes, and
then iterate five times to get more accurate values for $a_i$.
In this way $a_i$ is found from the mean particle size and the
particle separations, which are obtained directly from microscopy.

\begin{figure}
\begin{center}
\includegraphics[width=8cm]{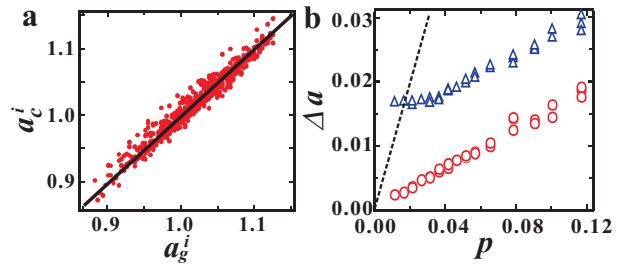}
\end{center}
\caption{(Color online) (a) Scatter plot of the calculated radius
$a_c^i$ from our method (Eqn.~\ref{Rioriginal}) as a
function of the given radius $a_g^i$ using data from a simulated
packing with polydispersity 7\%.
The solid line corresponds to $a_c^i = a_g^i$. 
(b) The particle size uncertainty $\Delta a$ found by analyzing simulation
data from packings with a given polydispersity, both without
noise (circles) and with noise added to the particle positions
(triangles).
The dashed line corresponds $\Delta a = p$. 
}
\label{sim-dis}
\end{figure}

To validate our method, we simulate polydisperse rcp samples using
the algorithm of Refs.~\cite{Xu,Ken}.  We use 512 particles with
mean radius $\bar{a}=1$ and polydispersity from 0.01 to 0.12,
generating 5 independent configurations for each polydispersity.
The particle size distribution is a Gaussian.  Using the simulated
position centers, we calculate the radii of the particles $a_c^i$
by our method.  Figure \ref{sim-dis}(a) shows a scatter plot of
$a_c^i$ as a function of the given radii $a_g^i$ from a simulation
with 7\% polydispersity.  The calculated radii are located around
$a_c^i = a_g^i$.  We define the uncertainty of the size estimation
as $\Delta a = \sqrt{\langle [(a_c^i - a_g^i)/a_g^i]^2 \rangle_i}$.
$\Delta a$ is plotted as a function of polydispersity $p$ as
circles in Fig.~\ref{sim-dis}(b).  We find $\Delta a \approx p/6$.
The polydispersity of $a_c^i$ matches that of $a_g^i$.

One experimental complication is that there is an uncertainty in the
position of each particle.  In our experiment, the uncertainties
are $0.024\bar{a}$ in $x$ and $y$ and $0.0395\bar{a}$ in $z$.
We add this positional uncertainty to the true simulated positions,
and then redetermine the particle radii.  As expected, this
increases the uncertainty $\Delta a$ of the final radii, shown
by the triangles in Fig.~\ref{sim-dis}(b).  $\Delta a$ increases
by $\sim 0.01$ compared to the case without positional noise.
Positional noise is fatal when the polydispersity is less than 0.02,
but otherwise our method results in more accurate radii even in
the presence of noise.

Next, we estimate each particle size of our experimental data
with our method.  Given that Eqn.~\ref{Rioriginal} requires
information about both a particle's nearest neighbors and also
second nearest neighbors, only particles sufficiently far from the
edges of our images have accurate sizes.  We modify our algorithm
slightly for the experimental data as follows.  We find the
coordination number $z_i$ of each particle, the number of
neighboring particles within a distance $2.8a$ (the first minimum
of the pair correlation function) \cite{KW}.  From the particles
in the interior of the sample, we find the average coordination
number $\bar{z} \approx 12$.  Then, for every particle, we estimate the
number of touching neighbors $T_i = 5 z_i/12$ where we round
$T_i$ to the nearest integer.  For particles at the edge of the
imaged volume, $T_i < 5$ as not all of the neighbors are imaged.
Then for each particle, when averages over contacting neighbors $j$
are done in Eqn.~\ref{Rioriginal}, these averages are over
the $T_i$ nearest neighbors.  After iterating Eqn.~\ref{Rioriginal}
to find all radii, the edge particles are removed by cropping the
data to a volume of 440 $\times$ 461 $\times$ 14.2 $\mu$m$^3$,
containing 217 816 particles.

\begin{figure}
\begin{center}
\includegraphics[width=6cm]{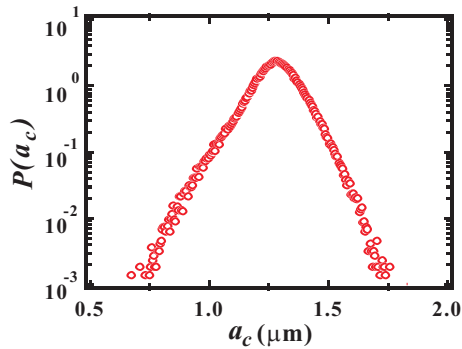}
\end{center}
\caption{(Color online) Probability of particle sizes in our
experimental sample.  The average size is 1.265 $\mu$m and
polydispersity is 6.7\%.
}
\label{exp-est}
\end{figure}

Based on these particles with their calculated sizes, the volume
fraction of this sample is found to be $\phi$ = 0.647 $\pm$ 0.007,
where the uncertainty of $\phi$ is due to the uncertainty in
determination of each particle size.  Figure \ref{exp-est} shows
a distribution of the estimated particle sizes.  This sample has
a polydispersity of 6.7\%.  Given this measured polydispersity,
Fig.~\ref{sim-dis}(b) shows that $\Delta a \approx 0.023$
(corresponding to $\bar{a} \Delta a=0.03$~$\mu$m).  The experimental
distribution is not a Gaussian and this is not an artifact of our
method, as a simulated Gaussian size distribution with positional
noise leads to a measured Gaussian size distribution.

\begin{figure}
\begin{center}
\includegraphics[width=7.3cm]{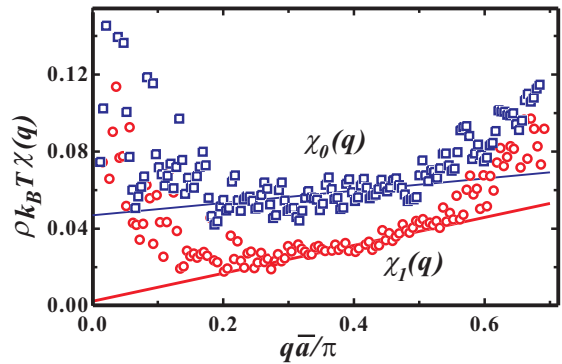}
\end{center}
\caption{
(Color online) $\rho k_B T \chi_0(q)$ (no approximation)
and $\rho k_B T \chi_1(q)$ (first order approximation of
Ref.~\cite{Berthier}), from
the experimental data.  Square symbols correspond to $\rho
k_B T \chi_0(q)$, which is proportional to $S(q)$ at small $q$.
Circle symbols correspond to $\rho k_B T \chi_1(q)$.  The lines
are linear fits to the data for $0.2 < q\bar{a}/\pi < 0.5$.
}
\label{chiq}
\end{figure}

Using our estimated particle sizes, we now study the wave
vector dependence of the compressibility $\chi_0 (q)$ and $\chi_1
(q)$ of our experimental data.  Figure \ref{chiq}
shows $\rho k_B T \chi_0(q)$ and $\rho k_B T \chi_1(q)$.
Our experimental data do not obey
periodic boundary conditions, and the effect of the boundaries
appears near $q = 0$.  $\chi_0(q)$ and $\chi_1(q)$
are independent of the choice of Fourier window functions for
$q\bar{a}/\pi >$ 0.2.  Thus we do a linear fit to $\rho k_B
T \chi_0(q)$ and $\rho k_B T \chi_1(q)$ in the region $0.2 <
q\bar{a}/\pi < 0.5$, shown as the lines in Fig.~\ref{chiq};
both functions have linear behavior in this region.  We find
$\rho k_B T \chi_1(0) = 0.002 \pm 0.004$, while $\rho k_B T
\chi_0(0) = 0.049 \pm 0.008$ as reported previously \cite{KW}.  The
uncertainties are due to the uncertainties of particle positions and
sizes, and the choice of the fitting range.  Our observation that
$\chi_1(q) \sim q$ shows that long wavelength density fluctuations
are suppressed.  This is consistent with the observations of
Berthier {\it et al.} and show that our system is 
incompressible and likely hyperuniform \cite{Berthier}.

Our data let us consider a new question, the relationship
between local environment and local ordering.  It is known that
crystallization occurs in samples with low polydispersity ($p <
0.08$) \cite{Schope,Henderson,Frenkel}.  However, crystal nucleation
is a microscopic phenomenon, that is, the crystal nuclei do not
necessarily ``know'' the bulk polydispersity.  We can use our
data to investigate the relationship between local ordering and
local polydispersity.

\begin{figure}
\begin{center}
\includegraphics[width=6cm]{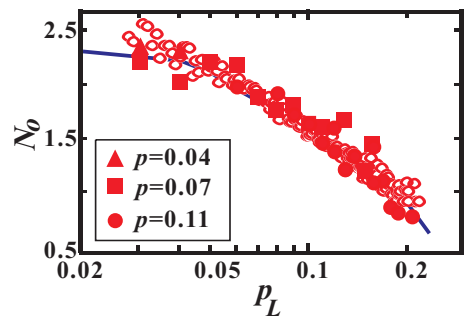}
\end{center}
\caption{(Color online) The number of ordered neighbors $N_o$ as
a function of the local polydispersity $p_{L}$ in the experiment
(open circles) and simulations (bulk polydispersities as
indicated in legend).  The line is the average of $N_o$
from 75 simulated systems with bulk polydispersity from 1\% to 12\%.
}
\label{crystal}
\end{figure}

We define the local polydispersity $p_{L}^i$ as 
\begin{equation}
p_{L}^i = \sqrt{\langle (a_n - a_i)^2 \rangle}/a_i
\label{localpoly}
\end{equation}
where $\langle a_n \rangle$ is the mean radius of the nearest
neighbor particles of particle $i$.  The nearest neighbors of
a particle are defined as those with centers separated by less
than 2.8$a$ \cite{KW}.  We calculate the bond order parameter
$d_6^{in}$ to quantify 
how the local structure compares between neighbors $i$
and $n$ \cite{Steinhardt, Gasser,Wolde}.  Two neighboring particles
are termed ``ordered neighbors" if $d_6^{in}$ exceeds a threshold
value of 0.5 \cite{Steinhardt,Gasser,Wolde}.  We then count the
number of ordered neighbors $N_o^i$ around each particle $i$.
$N_o^i = 0$ corresponds to random structure around particle $i$,
while $N_o^i > 7$ means that particle $i$ is in a crystalline
environment \cite{Wolde}.

Figure \ref{crystal} shows that the local polydispersity $p_L^i$
has a strong influence on local order $N_o^i$.  The open circles
show the result from our experiment.  Particles with low $p_L$
are more ordered than particles with high $p_L$:  that is, there
is a tendency for particles to order when the central particle size
$a_i$ is similar to its surrounding neighbors.  A similar result is
found from our simulated packings (closed symbols and solid line
in Fig.~\ref{crystal}), where the local polydispersity predicts
local order independent of the global polydispersity.  The agreement
between the simulations and the experiment is striking, especially
given that the simulation corresponds to an extremely fast quench,
whereas the experimental quench allows time for particles to
rearrange \cite{KW}.  Note that these conclusions are unchanged
when $a_i$ in Eqn.~\ref{localpoly} is replaced by $\langle a_n
\rangle$, although the trend shown in Fig.~\ref{crystal} is less
pronounced.  Our results are consistent with the prior knowledge
that polydispersity affects the ability to crystallize \cite{Schope,
Henderson, Frenkel}, but this is the first examination we are
aware of showing how polydispersity can have a local influence
on crystallization.  It suggests an intuitively reasonable idea,
that in a moderately polydisperse sample, crystalline nuclei are
more likely to form from locally monodisperse patches.

We note that the observed polydispersity of our sample (6.7\%)
helps explain a discrepancy we noted between
our observations \cite{KW} and those of Dullens {\it et al.}, who also
studied dense suspensions of sedimenting particles with similar
sedimentation rates \cite{Dullens}.  They observed that particles
formed crystals in all cases \cite{Dullens}, while our particles
pack randomly.  Their samples had a polydispersity of 5\%, while our
sample is 6.7\%.  Crystal nucleation is sensitive to polydispersity
in this range \cite{Frenkel} and this likely explains why our
sample avoids crystallization, and why the samples of Dullens {\it
et al.} crystallized.

To summarize, we have presented a method to estimate the sizes of
individual colloidal particles from experimental knowledge of
only their positions, and relying on the fact that the sample is
close-packed.  Numerical simulations confirm that our method is
robust even in the presence of realistic experimental noise.
Using the positions and sizes of over 200 000 random close packed
particles from our experiment, we confirm that our experimental
system is hyperuniform and incompressible.  Our results are
consistent with prior work \cite{Berthier} and the data can be
used with other algorithms for quantifying hyperuniformity in
polydisperse samples \cite{Zachary1}.  We also see a relationship
between local polydispersity and local order, confirming that
locally a higher polydispersity results in less ordered packing.

E.~R.~W.~was supported by a grant from the National Science
Foundation (CHE-0910707).

\end{document}